\documentclass[%
 reprint,
superscriptaddress,
 amsmath,amssymb,
 prl,
]{revtex4-2}

\usepackage[version=4]{mhchem}
\usepackage{graphicx}
\usepackage{dcolumn}
\usepackage{bm}

\usepackage{color}

\begin{document}

\preprint{APS/123-QED}

\title{High energy spin excitations in the quantum spin liquid candidate Zn-barlowite probed by resonant inelastic x-ray scattering}

\author{Rebecca W. Smaha}
\email{rsmaha@alumni.stanford.edu} 
\affiliation{Stanford Institute for Materials and Energy Sciences, SLAC National\\Accelerator Laboratory, Menlo Park, California 94025, USA}
\affiliation{Department of Chemistry, Stanford University, Stanford, California 94305, USA}
\altaffiliation[Present address: ]{Materials Science Center, National Renewable Energy Laboratory, Golden, Colorado 80401, USA}
\author{Jonathan Pelliciari}
\affiliation{National Synchrotron Light Source II, Brookhaven National Laboratory, Upton, NY, 11973, USA}
\author{Ignace Jarrige}
\affiliation{National Synchrotron Light Source II, Brookhaven National Laboratory, Upton, NY, 11973, USA}
\author{Valentina Bisogni}
\affiliation{National Synchrotron Light Source II, Brookhaven National Laboratory, Upton, NY, 11973, USA}
\author{Aaron T. Breidenbach}
 \affiliation{Stanford Institute for Materials and Energy Sciences, SLAC National\\Accelerator Laboratory, Menlo Park, California 94025, USA}
 \affiliation{Department of Physics, Stanford University, Stanford, California 94305, USA}
\author{Jack Mingde Jiang}
 \affiliation{Stanford Institute for Materials and Energy Sciences, SLAC National\\Accelerator Laboratory, Menlo Park, California 94025, USA}
 \affiliation{Department of Applied Physics, Stanford University, Stanford, California 94305, USA}
 \author{Jiajia Wen}
 \affiliation{Stanford Institute for Materials and Energy Sciences, SLAC National\\Accelerator Laboratory, Menlo Park, California 94025, USA}
\author{Hong-Chen Jiang}
 \affiliation{Stanford Institute for Materials and Energy Sciences, SLAC National\\Accelerator Laboratory, Menlo Park, California 94025, USA}
\author{Young S. Lee}
 \email{youngsl@stanford.edu}
 \affiliation{Stanford Institute for Materials and Energy Sciences, SLAC National\\Accelerator Laboratory, Menlo Park, California 94025, USA}
 \affiliation{Department of Applied Physics, Stanford University, Stanford, California 94305, USA}

\date{\today}

\begin{abstract}
A quantum spin liquid is a novel ground state that can support long-range entanglement between magnetic moments, resulting in exotic spin excitations involving fractionalized $S=\frac{1}{2}$ spinons. Here, we measure the excitations in single crystals of the spin liquid candidate Zn-barlowite using resonant inelastic X-ray scattering. By analyzing the incident polarization and temperature dependences, we deduce a clear magnetic scattering contribution forming a broad continuum that surprisingly extends up to $\sim$200 meV ($\sim$14$J$, where $J$ is the magnetic exchange). The excitation spectrum reveals that significant contributions arise from multiple pairs of spinons and/or antispinons at high energies.
\end{abstract}

\maketitle

A fascinating phenomenon in quantum magnets is the possibility of having magnetic excitations with fractional quantum numbers. In particular, this may occur in a quantum spin liquid (QSL), which is characterized by long-range quantum entanglement of the spins in the absence of long-range magnetic order.\cite{Balents2010,Norman2016,Broholm2020} Here, the fundamental $\Delta S=1$ excitations may fractionalize into pairs of $S=\frac{1}{2}$~spinons. A promising host of a QSL is the kagome lattice of corner-sharing triangles with magnetically frustrated antiferromagnetic spin-$\frac{1}{2}$ moments.\cite{Sachdev1992,Ran2007,Jiang2008,Yan2011,Depenbrock2012,Jiang2012,He2017,Mei2017,Liao2017,Zhang2020} A pre-eminent kagome QSL candidate is the mineral herbertsmithite \ce{(Cu3Zn(OH)6Cl2)}; however, a small percentage of excess \ce{Cu^2+} impurities are found on the nonmagnetic \ce{Zn^2+} sites between the kagome layers.\cite{Shores2005,Han2012,Fu2015,Han2016b} To better understand how  disorder and their coupling to the kagome moments affect the spin excitations, measurements on new materials are needed.

\begin{figure*}[t]
\includegraphics[width=\textwidth]{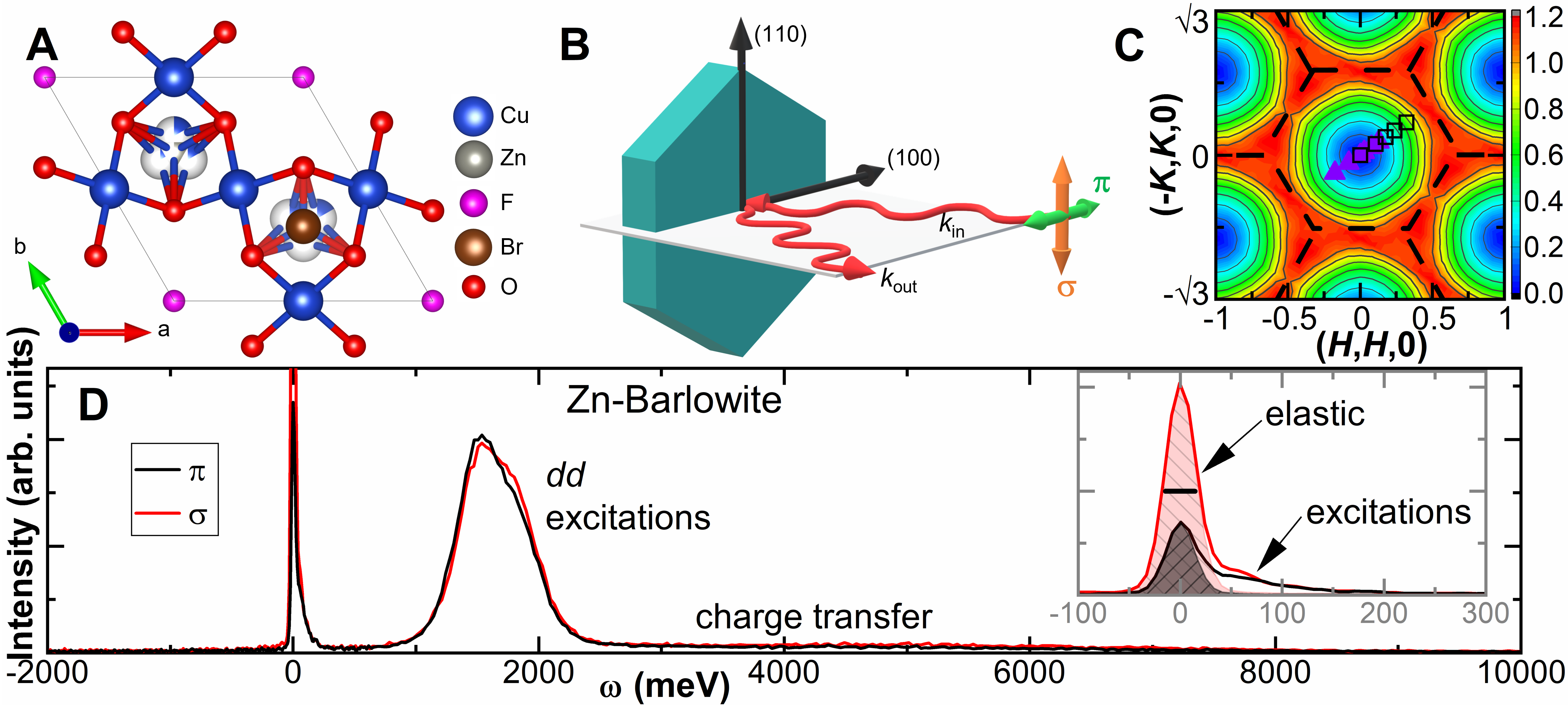}
\caption{\label{fig:fig1} A) Crystal structure of Zn-barlowite with $x=0.56$ (\textbf{Zn$_{0.56}$}) in the kagome (\textit{ab}) plane; H atoms are not shown, and the structure is from Ref. \cite{Smaha2020}. B) Schematic of the sample orientation on the beamline showing the ($H$0$L$) scattering plane and the $\pi$- or $\sigma$-polarization directions. C) Calculated static spin structure factor $S(\mathbf{q},S_z)$ with total spin $S_z=0$. The Brillouin zone is depicted with dashed lines, and the measured $q_{||}$ locations are shown as purple triangles and black squares for scattering angles $2\theta=90^{\circ}$ and $150^{\circ}$, respectively. D) RIXS signal of Zn-barlowite \textbf{Zn$_{0.56}$} at $(q_{||},q_\perp)=(0,1.0)$ collected in both $\pi$ and $\sigma$ polarizations. The band of $dd$ excitations is visible between $\sim$1--2.5 eV, and charge transfer scattering is visible as a broad hump between $\sim$3--6 eV. The inset magnifies the region near $\omega=0$ meV showing the elastic component and inelastic scattering. The black bar represents the instrumental resolution (29 meV). The elastic components were fit with Pseudo-Voigt lineshapes denoted by the shaded area (see Supplemental Material, Section IC). The data were collected at $T=30$ K and $2\theta=90^{\circ}$.}
\end{figure*}

A related spin-$\frac{1}{2}$ kagome material, Zn-substituted barlowite (\ce{Cu3Zn$_{x}$Cu$_{1-x}$(OH)6FBr}), was recently identified as a QSL candidate that has a different impurity environment and layer stacking arrangement compared to herbertsmithite.\cite{Feng2018,Smaha2018,Smaha2020,Smaha2020PRM}  We had synthesized the first single crystals of Zn-barlowite with a Zn substitution of $x=0.56$, referred to as \textbf{Zn$_{0.56}$}.\cite{Smaha2020} Its hexagonal structure is shown in Fig. \ref{fig:fig1}A: the kagome layers are fully occupied by \ce{Cu^2+}, while the interlayers have a majority of nonmagnetic \ce{Zn^2+} ions interspersed with \ce{Cu^2+} impurities.\cite{Smaha2020PRM} \textbf{Zn$_{0.56}$} has no long-range magnetic order down to $T=0.1$ K, making it a good QSL candidate.\cite{Smaha2020} However, the lack of large single crystals have impeded measurements of the spinon physics with neutron scattering, a technique which provides high energy resolution and a well-defined cross-section for scattering from $\Delta S=1$ excitations. 

Resonant inelastic X-ray scattering (RIXS) offers a unique opportunity to study this kagome QSL candidate. RIXS is a powerful probe of elementary electronic excitations, including magnetic excitations, and has the capability of measuring small single crystals ($<1$ mm in length).\cite{Ament2011,Schlappa2018,Pelliciari2021,Pelliciari2021FeSe} Furthermore, RIXS naturally probes significantly higher energy transfers than neutrons and has a different cross-section for the inelastic scattering. For magnetic scattering, the RIXS cross section may detect $\Delta  S=0$ as well as $\Delta S=1$ excitations, and therefore can probe excitations that evade the sensitivity of neutrons.\cite{Schlappa2018,Nag2020} While the application of RIXS to materials with low-energy or overlapping excitations has been limited by the resolution, significant advances have recently improved the achievable resolution to 15--30 meV for soft X-rays. Recent experimental and computational advances in soft X-ray RIXS have focused on compounds with 1D chains or 2D square lattices of \ce{Cu^2+}.\cite{Ament2009,Braicovich2010,Haverkort2010,Peng2018,Schlappa2018} However, even though the kagome quantum magnets with \ce{Cu^2+} cations display much interesting physics, no experimental RIXS investigation has yet been published.

We present RIXS measurements on small single crystals of the kagome QSL candidate Zn-barlowite \textbf{Zn$_{0.56}$}.\cite{Smaha2020} RIXS data were collected on beamline 2-ID at the NSLS II at Brookhaven National Laboratory at the Cu $L_3$-edge ($\approx$930 eV) (Fig. \ref{fig:fig1}B).\cite{Dvorak2016,Lebert2020}  The energy resolution, shown as the black bar in Fig. \ref{fig:fig1}D(inset), was approximately 29 meV. X-ray absorption spectroscopy confirms the presence of \ce{Cu^2+} (Fig. S1 in the Supplemental Material). We collected RIXS data at a scattering angle of $2\theta=90^{\circ}$ in both $\pi$ and $\sigma$ incoming photon polarizations. At $2\theta=90^{\circ}$ with $\pi$ polarization, the elastic charge scattering and phonon scattering should be suppressed relative to the magnetic scattering. However, precise quantification may be complicated by the lower point group symmetry of the kagome \ce{Cu^2+} $d$-orbitals compared to square lattice cuprates.\cite{Haverkort2010PRB}  Figure \ref{fig:fig1}D shows a wide energy loss ($\omega$) range of the data, where scattering from the $dd$ excitations is visible between 1--2.5 eV, and charge transfer scattering is present as a broad, weak peak centered at $\sim$4 eV. The inset, which magnifies the region near $\omega=0$ meV, shows the suppression of the elastic line in $\pi$ polarization compared to $\sigma$ polarization. All data were normalized by the area of the $dd$ excitations. After fitting the elastic component with a Pseudo-Voigt (shaded in the inset), we observed inelastic scattering extending up to approximately 200 meV.

\begin{figure*}
\includegraphics[width=\textwidth]{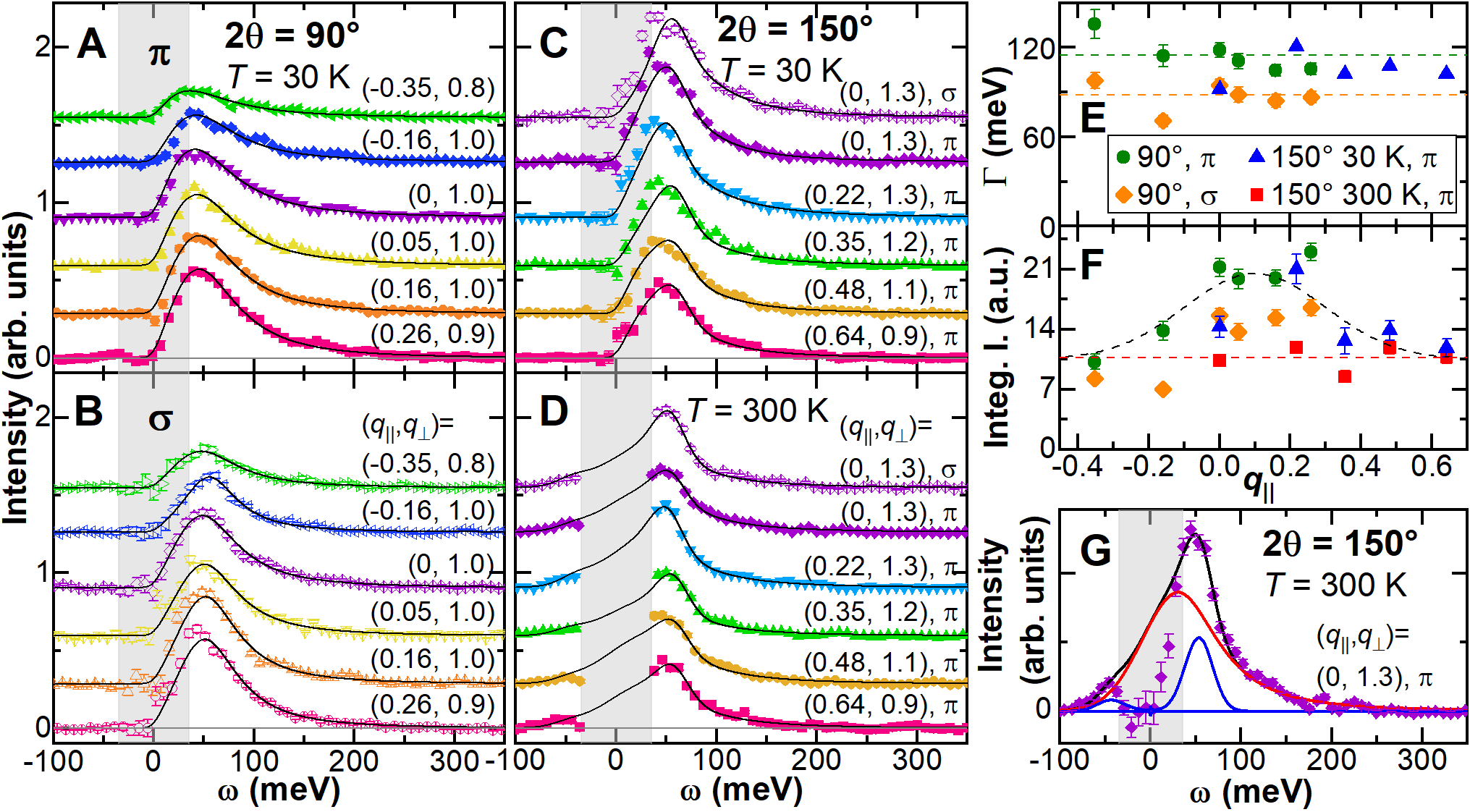}
\caption{\label{fig:fig2} A) and B) The $q$-dependence of the inelastic scattering of Zn-barlowite \textbf{Zn$_{0.56}$} collected at $T=30$ K and scattering angle $2\theta=90^{\circ}$ in both A) $\pi$ and B) $\sigma$ polarizations. C) and D) The $q$-dependence of the inelastic scattering of \textbf{Zn$_{0.56}$} collected at C) $T=30$ K and D) $T=300$ K at $2\theta=150^{\circ}$. Both $\sigma$ (open symbols) and  $\pi$ (closed symbols) polarization are shown. In A)--B), the data were fit with a damped harmonic oscillator (DHO) curve from -100--500 meV with the DHO center fixed at 65 meV. In C)--D), the data were fit with a DHO and a phonon mode (see Supplemental Material, Section IE) from -100--500 meV, masking out -35--35 meV. The best fits are shown as black lines. The shading denotes in A)--B) a region of high uncertainty due to the elastic subtraction and in C)--D) these region were masked out during fitting. Vertical offsets were applied to separate the spectra. E)--F) The fitted DHO width $\Gamma$ and integrated intensity as a function of $q_{||}$. Horizontal dashed lines denote average values, while curves are guides to the eye. G) An example of the ``DHO + phonon" model showing the fit components (red and blue lines, respectively) and the overall fit (black line).}
\end{figure*}

For this quasi-2D material, $\vec{q}$ can be decomposed into in-plane and out-of-plane components $(q_{||},q_\perp)$. Here, $q_{||}$ is along ($H$,0,0), and $q_\perp$ is along (0,0,$L$) (see Fig. S2). For weakly coupled layers, the scattering at high energies should depend primarily on $q_{||}$ within the kagome layer and be weakly dependent on (or independent of) $q_\perp$. Subtracting the elastic component reveals the inelastic scattering as a broad continuum (Fig. \ref{fig:fig2}A--B). The lack of any sharp components suggests that scattering from individual phonon modes is weak under these experimental conditions,\cite{Ament2011} although there may be some contribution within the energy range of the continuum (Fig. S6). We fit the inelastic scattering with a damped harmonic oscillator (DHO) cross section (see Supplemental Material, Section ID), which provides a qualitatively good description of the data. The fitted width ($\Gamma$) and integrated intensity are weakly dependent on $q_{||}$ and are plotted in Fig. \ref{fig:fig2}E--F, respectively. The $\pi$-polarized data, which should be less affected by the charge scattering, has a consistently higher integrated intensity than the $\sigma$-polarized data, indicating that the continuum likely has a substantial magnetic component.

To measure larger wavevectors, we also collected data at $2\theta=150^{\circ}$ (Fig. \ref{fig:fig2}C--D). The inelastic scattering at $T=30$ K has a broad tail that extends up to $\sim$200 meV, consistent with that at $2\theta=90^{\circ}$. At $T=300$ K, the high energy scattering is somewhat reduced, revealing a more well-defined peak near 60 meV that is likely due to phonons, as this energy is consistent with previously calculated phonon modes \cite{Fu2021}. We therefore fit the $2\theta=150^{\circ}$ data at both temperatures with a model consisting of a DHO and a phonon mode (see Supplemental Material, Sections IE and II). As seen in the individual fit components drawn in Fig. \ref{fig:fig2}G, when the phonon mode is taken into account, a DHO with similar parameters as previously determined describes the remaining continuum. The integrated intensities (Fig. \ref{fig:fig2}F) of the $T=30$ K data have a weak maximum at small $q_{||}$ values. In contrast, the room temperature intensities are relatively suppressed and are flat across $q_{||}$. Thus, if the broad continuum is magnetic in origin, then the spin correlations grow upon cooling as would generally be expected.

\begin{figure}
\includegraphics{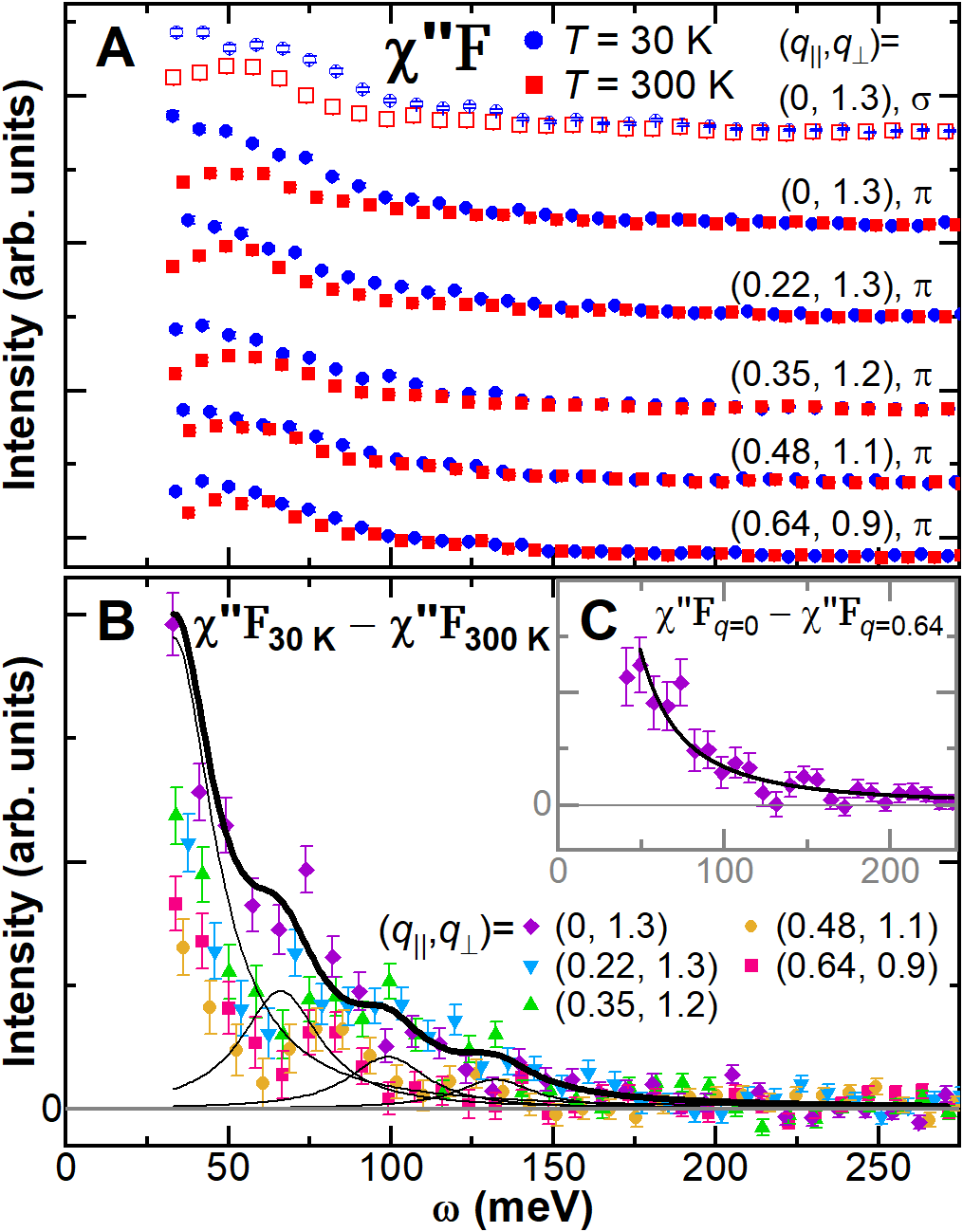}
\caption{\label{fig:fig3} A) The $q$-dependence of the susceptibility times the form factor ($\chi''F$) of Zn-barlowite \textbf{Zn$_{0.56}$} at $T=30$ K and room temperature. Vertical offsets were applied to separate the spectra. B) Enhancement of $\chi''F$ at low temperature and $\pi$ polarization, expressed by subtracting the $T=300$ K data from the $T=30$ K data (i.e., $\Delta\chi''F$). A series of Lorentzians 
representing single and multiple pairs of spinon excitations (thin lines) and their sum (thick line) are overlaid. C) The $q_{||}$-dependent subtraction $\chi''F_{q_{||}=0}-\chi''F_{q_{||}=-0.64}$ $\chi''F$ at $T=30$ K. The line is the result of fitting to a power law $\sim \frac{1}{\omega^{2}}$. All data were collected at $2\theta=150^{\circ}$.}
\end{figure}

To further investigate the magnetic origin of the signal, we divided the total inelastic scattering by the thermal factor ($n(\omega)+1$). Hence, if the inelastic scattering is proportional to $S(q,w)F(q,w)$, where $F(q,w)$ is a temperature-independent form factor, then the phonon ``background'' contribution to the resultant imaginary part of the general susceptibility, $\chi''$, should not change much between these temperatures (see Supplemental Material, Section V). Therefore the change in $\chi''F$ between high and low temperature should be mostly magnetic in origin, and this comparison is plotted in Fig. \ref{fig:fig3}A. Again, the signal is clearly stronger at low temperatures, indicating enhanced spin correlations. (If spin-phonon interactions are relevant, the phonon dynamics may also evolve with the spin correlations.) A larger difference is seen at small $q_{||}$, where the quantity ($\chi''F_{T=30 K}-\chi''F_{T=300 K}$) is shown in Fig. \ref{fig:fig3}B. Figure \ref{fig:fig3}C shows the $q_{||}$-dependent subtraction $\chi''F_{q_{||}=0}-\chi''F_{q_{||}=-0.64}$ at $T=30$ K, revealing a weak $q_{||}$-dependence to the low temperature scattering.

Thus, both the polarization and temperature dependences of the inelastic scattering suggest the presence of a magnetic continuum up to $\sim$200 meV ($\sim 14 J$). The spectrum may be empirically described by an overdamped harmonic oscillator (with $\Gamma \approx 110$ meV $>$ $\omega_0 \approx 65$ meV). Due to the instrumental resolution of 29 meV, we cannot accurately determine any possible power-law or exponential dependence of the intensity near $\omega=0$ meV. The RIXS data cannot resolve the question of gapless versus gapped spin excitations at very low energies ($\ll J \sim 14$~meV) for this kagome QSL material. 

Now, because \textbf{Zn$_{0.56}$} does not have long-range magnetic order at any measured temperature, the broad continuum does not originate from scattering from conventional magnons. This leads to an interpretation in terms of scattering from pairs of spinons or spinon-antispinon pairs.\cite{Schlappa2018,Ko2010} The fact that the signal extends to a high energy of $\sim$14$J$ suggests that the entirety of the scattering does not arise from a single pair of spinon excitations. The high-energy tail of the RIXS scattering can be empirically fit to a power law $1/\omega^{\alpha}$ with $\alpha\simeq2$, as shown in Fig. \ref{fig:fig3}C. In the context of scattering from a single spinon pair, we fit the magnetic signal in Fig. \ref{fig:fig3}C with the functional form $S(q,\omega)=A/\omega^{2-\eta}$ used to describe possible QSL states.\cite{Hermele2008,Chubukov1994,Coldea2003} The fitted $\eta$ would have to be quite small ($<0.25$) to describe the continuum within these scenarios.

Therefore, the high energy RIXS scattering in Zn-barlowite \textbf{Zn$_{0.56}$} likely includes significant contributions from multiple spinon pairs. Since RIXS can measure spin excitations with $\Delta S =0$, scattering from spinon-antispinon pairs may also be present. Calculations for the Raman cross section of the Dirac spin liquid on the kagome lattice are instructive: Ko et al. predicted a broad spectrum of scattering extending up to $\sim$100 meV for herbertsmithite at $q=0$, which includes significant contributions from up to three spinon-antispinon pairs.\cite{Ko2010} Here, to describe scattering from multiple pairs of spinons and/or antispinons, we modeled the data in Fig. \ref{fig:fig3}B with four Lorentzian peaks representing 1-pair, 2-pair, 3-pair, and 4-pair scattering. We find that the intensity of the Lorentzians decrease as 1/$\omega_0^2$, where $\omega_0$ denotes the center of each peak. The position of the 1-pair peak occurs at $\sim$35 meV, close to the peak position seen in magnetic Raman measurements.\cite{Wulferding2010,Fu2021} Interestingly, the intensity of the 1-pair peak is roughly 70\% of the total spectral weight, similar to the result for the spin-1/2 chain QSL.\cite{Karbach1997,Mourigal2013}

The enhanced scattering at $q_{||}=0$ further supports the scattering originating from more than one spinon pair. We calculated the static spin structure factor using density matrix renormalization group techniques \cite{White1992} on the $S=\frac{1}{2}$ antiferromagnetic Heisenberg model on the kagome lattice (Figs. \ref{fig:fig1}C and S13). The scattering, which represents the energy-integrated $\Delta S=1$~scattering, is very weak at $q_{||}=0$ as expected for antiferromagnetic correlations and is not consistent with our data. However, the signal at small $q_{||}$ may be enhanced when considering inelastic scattering from multiple pairs of spinons. We inspected the difference of the spin structure factors  $dS(\mathbf{q},S_z)=S(\mathbf{q},S_z)-S(\mathbf{q},0)$, where $S_z$ denotes the number of pairs of spinon-spinon excitations, and find improved consistency between $dS(\mathbf{q},2)$ and the RIXS results (see Fig.~S13). Moreover, the signal at $q_{||}=0$ may have a substantial contribution from scattering from spinon-antispinon pairs (as discussed above regarding the Raman predictions\cite{Ko2010}); however, the RIXS cross section for these processes has not yet been calculated.

\begin{figure}
\includegraphics{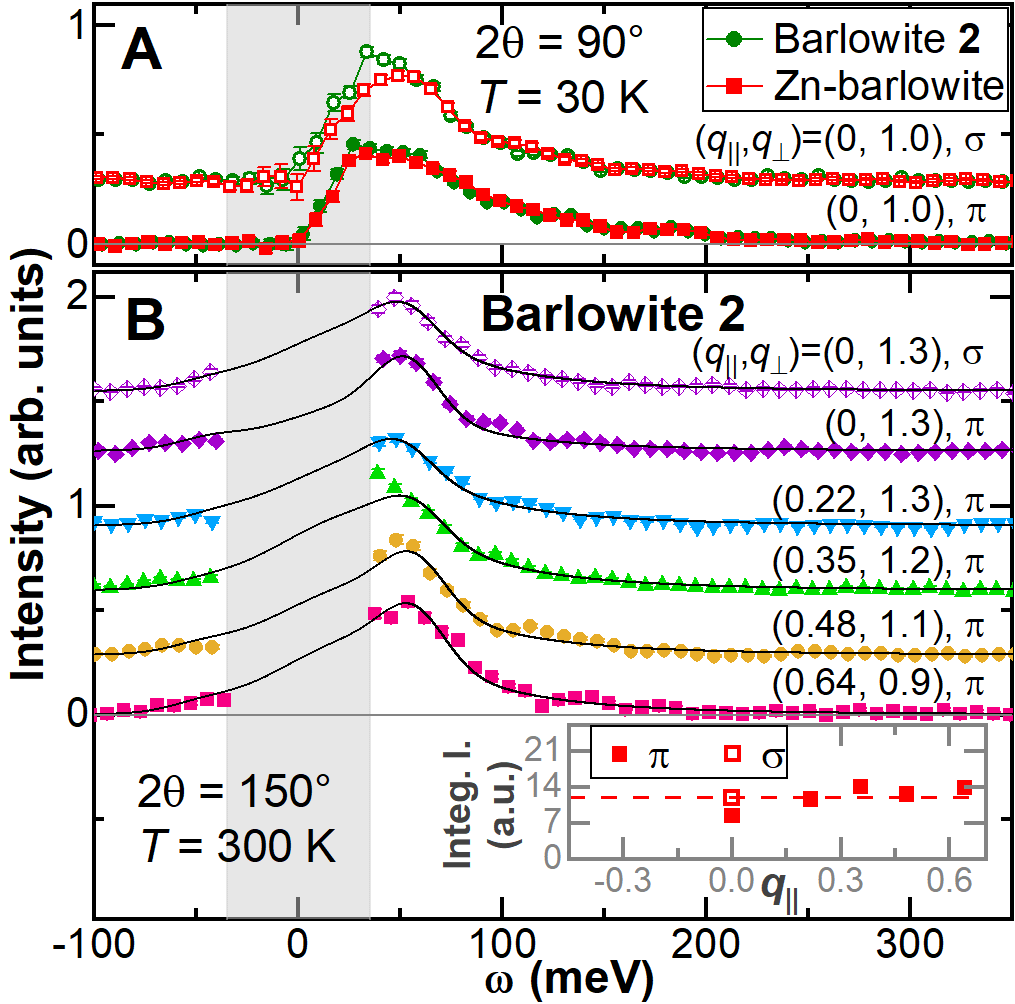}
\caption{\label{fig:fig4} A) Comparing the inelastic scattering of Zn-barlowite \textbf{Zn$_{0.56}$} and barlowite \textbf{2} collected at $q_{||}=0$, $T=30$ K, and $2\theta=90^{\circ}$. The shading denotes a region of high uncertainty due to the elastic subtraction. B) The $q$-dependence of the inelastic scattering of barlowite \textbf{2} collected at $T=300$ K and $2\theta=150^{\circ}$. The data were fit with the ``DHO + phonon" model (Supplemental Material, Section VI); the best fits are shown as black lines. The shaded region was masked out during fitting. Vertical offsets were applied to separate the spectra. Inset: fitted integrated intensity as a function of $q_{||}$. The horizontal dashed line is the average value.}
\end{figure}

We additionally collected RIXS data on small crystals of the magnetically ordered parent compound \ce{Cu4(OH)6FBr}, barlowite \textbf{2}, which does not have disorder related to cation substitution.\cite{Smaha2018,Smaha2020} The parent compound has $T_N\sim 10$~K and maintains hexagonal symmetry at low temperatures,\cite{Smaha2020} which is distinct from the variant that becomes orthorhombic.\cite{Han2014,Jeschke2015,Pasco2018,Feng2018,Henderson2019,Smaha2018,Smaha2020} Both \textbf{Zn$_{0.56}$} and barlowite \textbf{2} exhibit qualitatively similar RIXS spectra at low temperatures (Fig. \ref{fig:fig4}A). Data collected at $T=300$ K (Fig. \ref{fig:fig4}B and inset) likewise appear similar to Zn-barlowite, exhibiting a broad continuum. 

The similarity of the inelastic spectrum in barlowite 2 and Zn-barlowite indicates that spinon-based scattering can dominate at high energies even if magnetic order occurs at low temperatures. Furthermore, this shows that the high energy dynamics are not sensitive to interlayer disorder or interactions at much smaller energy scales, which eventually play a role in the ground state physics.\cite{Smaha2020} In other systems, RIXS measurements on a $S=\frac{1}{2}$ chain compound that orders at low temperatures revealed the presence of 4-spinon scattering.\cite{Schlappa2018} Even in two dimensions, the ordered $S=\frac{1}{2}$ square lattice antiferromagnet shows evidence for spinon continuum excitations at high energies probed with neutrons.\cite{Piazza2015} Future studies of Zn-barlowite and barlowite at the oxygen $K$-edge will be useful to better understand the multi-spinon excitations.\cite{Schlappa2018,Xiong2020}

Our data reveal that the magnetic excitations in the QSL candidate Zn-substituted barlowite forms a broad continuum extending up to $\sim$14$J$. The results indicate scattering from multiple spinon-antispinon and/or spinon-spinon pairs---a remarkable first sighting for the RIXS technique. Further work on understanding the multi-spinon excitations for the kagome QSL's (and their competing states) and calculating the appropriate RIXS cross-sections would be most illuminating. 

\begin{acknowledgments}
The work at Stanford and SLAC in the Stanford Institute for Materials and Energy Sciences (SIMES) was supported by the U.S. Department of Energy (DOE), Office of Science, Basic Energy Sciences (BES), Materials Sciences and Engineering Division, under Contract No. DE-AC02-76SF00515. This research used beamline 2-ID of the National Synchrotron Light Source II, a U.S. Department of Energy (DOE) Office of Science User Facility operated for the DOE Office of Science by Brookhaven National Laboratory under Contract No. DE-SC0012704. Part of this work was supported by the Laboratory Directed Research and Development project of Brookhaven National Laboratory No. 21-037. Part of this work was performed at the Stanford Nano Shared Facilities (SNSF), supported by the National Science Foundation under award ECCS-2026822. R.W.S. was supported by the Department of Defense through the NDSEG Fellowship Program and by a NSF Graduate Research Fellowship (DGE-1656518). We thank Senthil Todadri and Yao Wang for helpful discussions and A.F.LaFranchi for assistance with Python.
\end{acknowledgments}



%

\end{document}